# Evapotranspiration trends over the last 300 years reconstructed from historical weather station observations via machine learning

Haiyang Shi

Department of Civil and Environmental Engineering, University of Illinois at Urbana-Champaign, Urbana, IL 61801, USA

Correspondence: haiyang@illinois.edu; haiyang.shi9473@outlook.com

## Abstract

Estimating historical evapotranspiration (ET) is essential for understanding the effects of climate change and human activities on the water cycle. This study used historical weather station data to reconstruct ET trends over the past 300 years with machine learning. A Random Forest model, trained on FLUXNET2015 flux stations' monthly data using precipitation, temperature, aridity index, and rooting depth as predictors, achieved an $R^2$ of 0.66 and a KGE of 0.76 through 10-fold cross-validation. Applied to 5267 weather stations, the model produced monthly ET data showing a general increase in global ET from 1700 to the present, with a notable acceleration after 1900 due to warming. Regional differences were observed, with higher ET increases in mid-to-high latitudes of the Northern Hemisphere and decreases in some mid-to-low latitudes and the Southern Hemisphere. In drylands, ET and temperature were weakly correlated, while in humid areas, the correlation was much higher. The correlation between ET and precipitation has remained stable over the centuries. This study extends the ET data time span, providing valuable insights into long-term historical ET trends and their drivers, aiding in reassessing the impact of historical climate change and human activities on the water cycle and supporting future climate adaptation strategies.

**Introduction**

Historical estimates of evapotranspiration (ET) are crucial for understanding the impact of climate change and human activities on the water cycle [1,2]. By developing and analyzing historical ET data, we can reveal how long-term climate changes have influenced the global water cycle and gain insights into how ecosystems responded to climate variations in historical periods.

However, there are several limitations to estimating historical ET. Firstly, before 1970, there was a significant lack of satellite remote sensing observations related to vegetation, climate, and land water cycles. This makes it difficult to apply many ET estimation models, which often rely on satellite data [3–6]. Additionally, historical ground observation data (such as meteorological data) have limited spatial and temporal coverage, especially for early records, making it challenging to assess global ET from decades or centuries ago accurately.

Fortunately, recent developments and reprocessing of historical meteorological observation data [7] have greatly improved the availability of crucial meteorological data for estimating historical ET. This advancement, combined with modern flux tower observations [8] and meteorological data, enhances the possibility of reconstructing historical ET. Previous studies have shown that using machine learning (integrate flux observations and meteorological data) [9–11] or boundary layer theory [12] can estimate ET at weather station scales with relatively high accuracy. By applying these new methods to recently released historical weather station data, it is now possible to estimate the dynamic changes in ET over the past few centuries.

Therefore, this study combines modern flux and meteorological observations to develop a site-scale ET estimation model using machine learning based on a few meteorological variables (temperature and precipitation). This model is then applied to historical weather stations. Monthly ET data from 5267 weather stations spanning over 300 years are generated, and the long-term trends and driving factors of ET are analyzed. We significantly extended the current ET data's time span, offering valuable insights into how historical climate changes (including pre-industrial periods) have impacted the water cycle. It also provides more reliable scientific evidence for future water resource management and climate adaptation strategies.

**Results**

The monthly ET estimation model achieved high accuracy during 10-fold cross-validation at flux tower sites, with an $R^2$ of 0.66 and a KGE of 0.76 (Supplementary Fig. 1). The historical weather station ET data produced by this model show that most station records start before 1900 (Fig. 1), with the earliest records estimating ET for all 12 months even before 1700. The number of stations increased significantly after 1850, and many stations have records spanning over 100 years. Most of these stations (Fig. 2) are located in North America, Western Europe, and Australia, with sparse coverage in China, Africa, and South America. Before 1800, most stations were in Western Europe (Fig. 2). After 1850, the station density in North America, Australia, and India also increased significantly (Supplementary Fig. 2). The estimated annual ET values (Supplementary Fig. 3) are relatively low compared to some state-of-the-art ET estimates for the same regions after 2000 [5]. It is possibly because the historical weather stations were surrounded by vegetation with generally low coverage, making them less representative of land cover types with higher ET, such as forests and croplands.

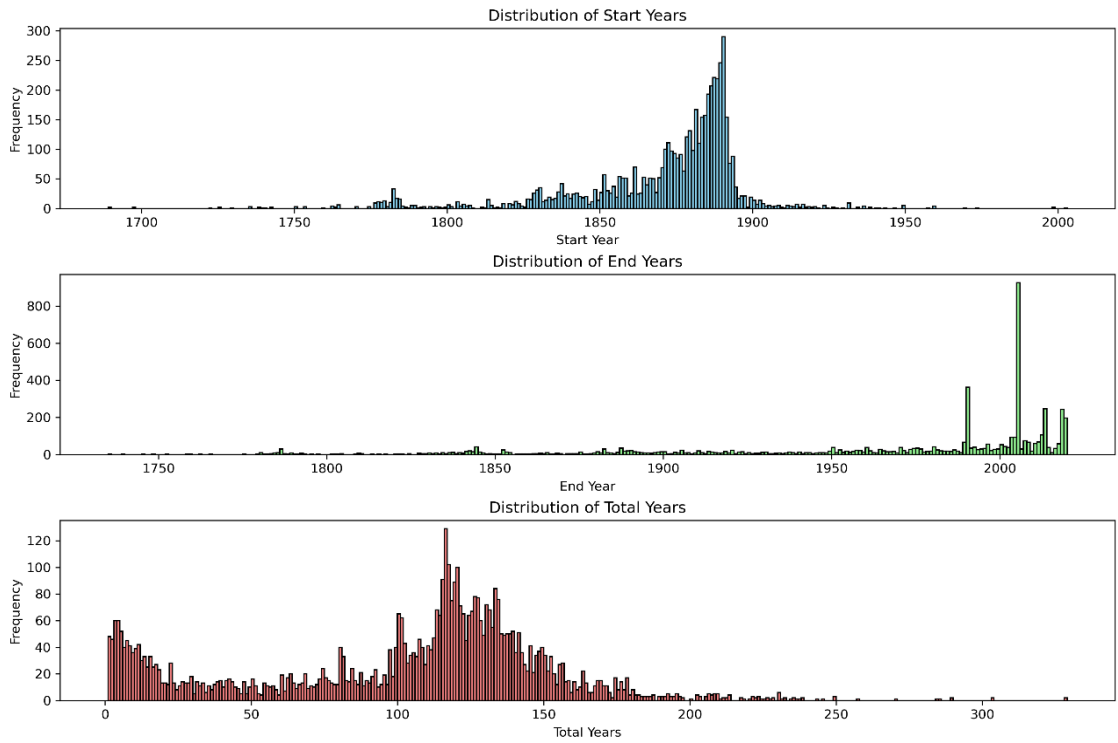

Fig. 1 Distribution of start year, end year, and total number of years of the produced weather station ET datasets.

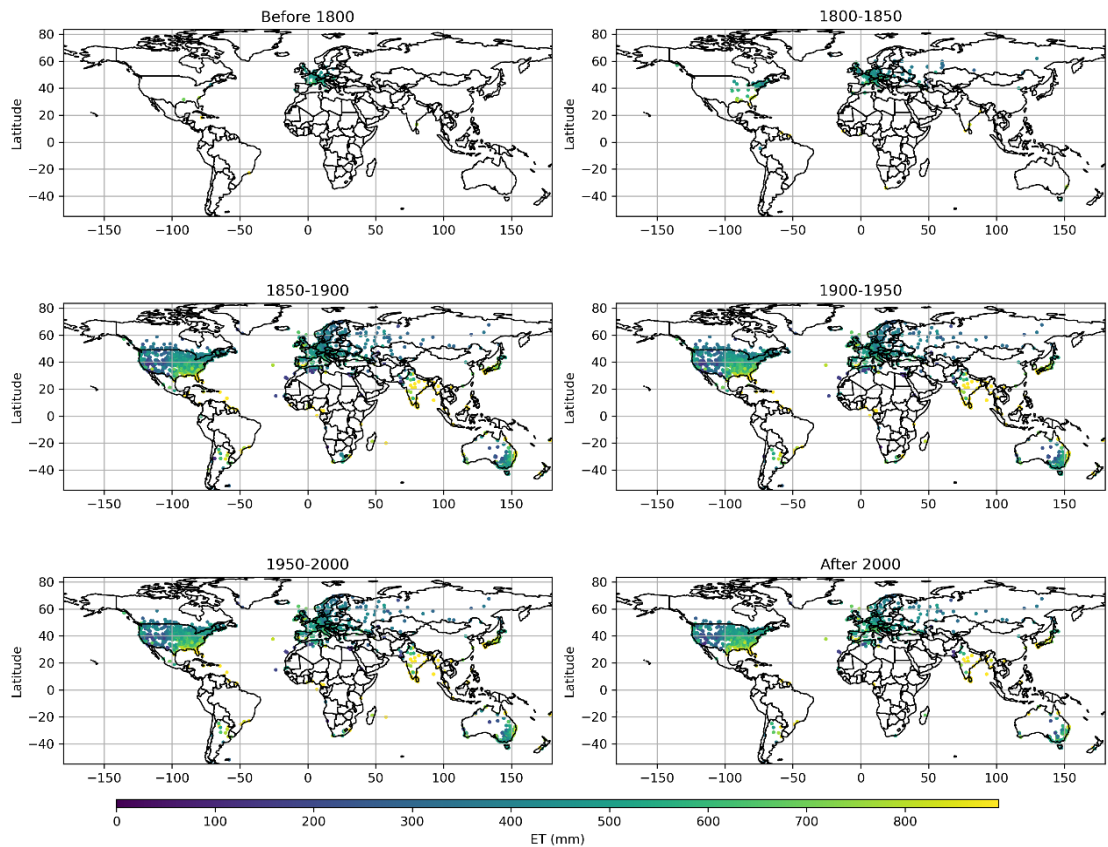

Fig. 2 Annual ET values of weather stations at different time periods.

Historical ET data shows that global ET has been increasing from 1700 to the present (Fig. 3). The interannual trend before 1850 might be more uncertain due to the limited number of stations. The fluctuation in the average ET trend significantly decreased after 1850, likely because the increase in stations and data reduced the variability of the median values. Interestingly, the first 50 years after 1839 showed a decline in ET (-2.28 mm) compared to the period from 1789 to 1839. However, after 1889, the rate of ET increase accelerated significantly, particularly during the period from 1889 to 1939, which saw an increase of 20.56 mm. Given that the ET values in this study are relatively low, this increase could represent almost 4%. This suggests that the rapid warming since the Industrial Revolution has led to a significant rise in ET, especially from 1889 to 1939. The rapid rise in $CO_2$ concentrations might also have contributed to increased ET through vegetation greening, but since the vegetation around weather stations typically has low coverage, the fertilization effect of $CO_2$ on ET remains uncertain. From 1939 to 1989, the rate of ET increase slowed significantly, with an increase of only 4.44 mm, which correlates with a slower rate of temperature increase during this period, especially with the temperature decline in the years following 1940 [13]. In the last thirty years, however, the rate of ET increase has risen again (+14.26 mm). Regarding regional differences (Fig. 3) in ET changes and rates, most stations in the mid-to-high latitudes of the Northern Hemisphere (above 40 degrees latitude) show an increase in ET. Some areas, such as central Canada, have higher increase rates. However, in the mid-to-low latitudes of the Northern Hemisphere and in the Southern Hemisphere, the trend of increasing ET is not as evident, with many stations showing a decrease in ET (e.g., Australia and the southwestern United States).

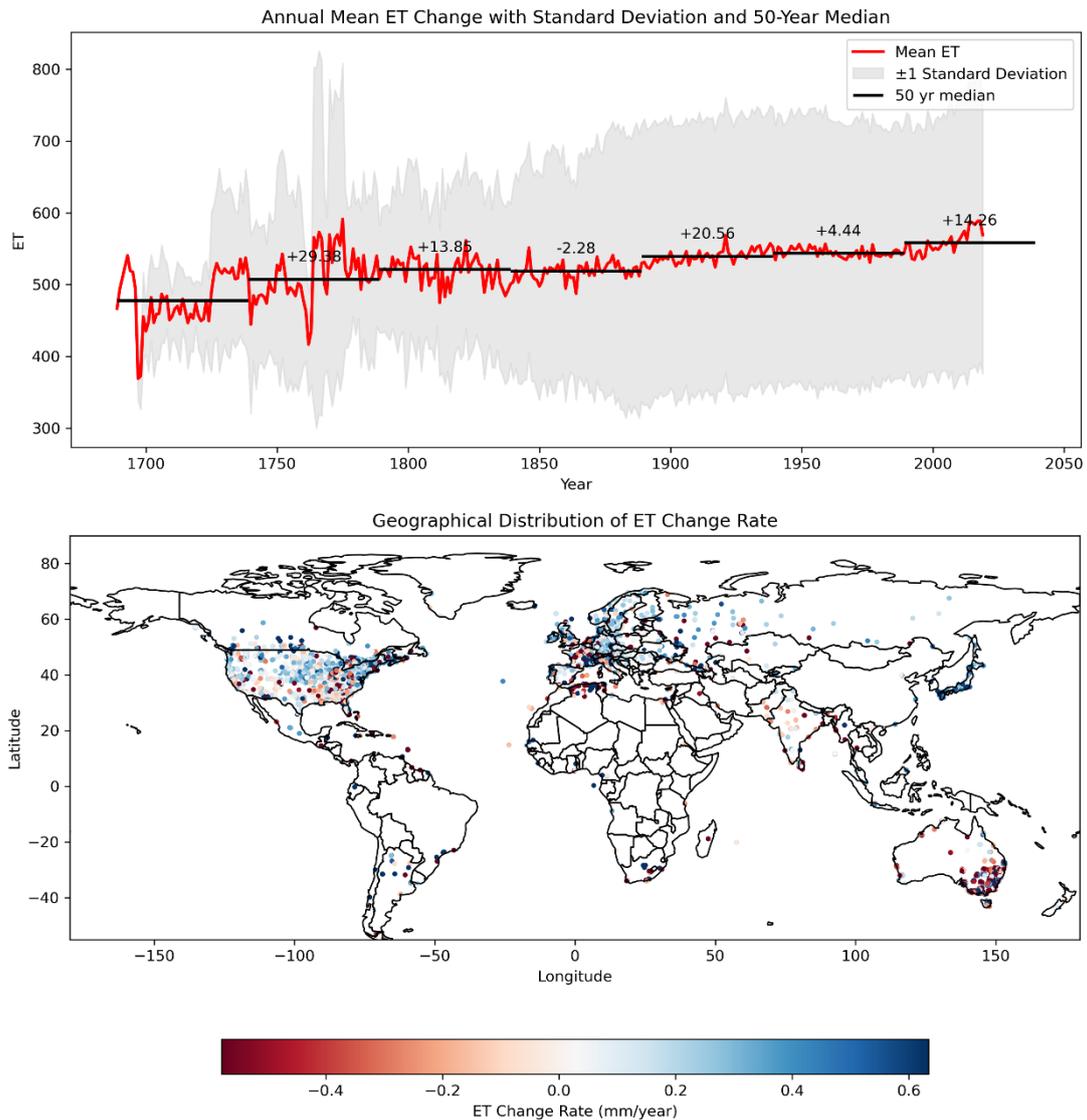

Fig. 3 Rates of ET trends at weather stations and their geographic variation over the last 300 years.

Before 1850, ET data mainly came from stations in Western Europe, so the correlation between ET, precipitation, and temperature might have been quite different compared to post-1850 data. After 1850, the interannual dynamics of ET and its correlations with precipitation and temperature remained relatively consistent, with correlation coefficients showing stability over different periods. Post-2000, the correlation between ET and temperature (Ta) in drylands slightly increased from 0.35 to 0.38. Globally, the correlation between ET and both temperature and precipitation has been around 0.6. However, in drylands, the correlation between ET and temperature is lower (ranging from 0.33 to 0.38). In humid areas, the correlation is much higher, between 0.83 and 0.85, indicating that temperature increases have predominantly driven ET increases in these regions. Interestingly, the correlation between ET and precipitation in water-limited drylands did not show a higher value, nor did it show a lower value in less water-limited humid areas. This suggests that on a 50-year analysis scale, the interannual variability [14] of precipitation's impact on ET has not significantly changed from 1850 to the present.

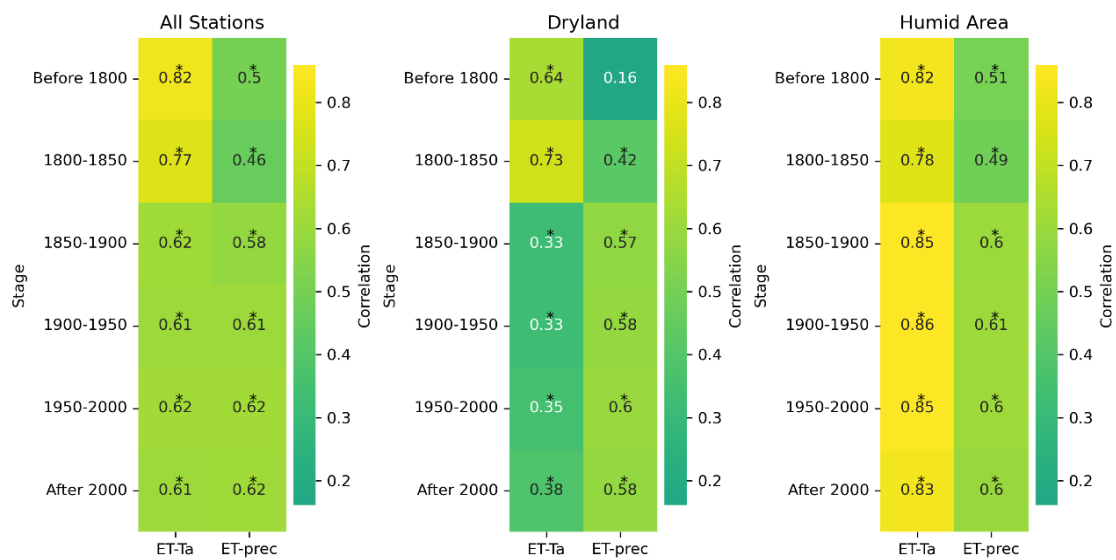

Fig. 4 Correlation of annual ET at weather stations with precipitation and temperature dynamics at different historical periods and in drylands and humid areas. The note '*' illustrates significant levels of $p < 0.05$.

**Discussion**

Previous studies on ET trends have relied heavily on satellite remote sensing [3–5,15,16], which only allows for the analysis of trends since the 1980s. Other methods based on climate models or earth system models [17–19] have significant uncertainties and limitations in historical periods with sparse observations. This study makes full use of historical weather station data to estimate ET, enabling the analysis and attribution of ET trends over the past 300 years. This addresses the gap in our understanding of ET changes before and after the Industrial Revolution.

However, the study has some limitations, mainly due to the representativeness of weather stations for the global surface and uncertainties in other data. Historical weather stations are scarce in high ET areas such as tropical rainforests, and stations are less likely to be installed in areas with high vegetation cover. This makes it challenging to represent the response of different ecosystems (especially forests) to long-term climate change. Additionally, the practice of using nearby stations to fill in missing precipitation or temperature data (see Methods) may introduce uncertainties. Temperature is generally more consistent between nearby stations compared to precipitation, which is often more spatially heterogeneous. However, since the model used in this study operates on a monthly scale rather than a daily scale, this error may be reduced. Most stations in this study are located in humid areas where ET is less limited by water availability, thus reducing the likelihood of significant prediction errors due to precipitation data uncertainties. Future research could benefit from reconstructing historical data for variables like VPD and soil moisture to further improve prediction accuracy, as many process-based ET models do not rely directly on precipitation data [5,6,15]. Additionally, global-scale historical or reconstructed data on land cover changes [17,20], vegetation [21], nitrogen deposition, aerosol concentrations, and $CO_2$ levels are sparse and uncertain. This limits the attribution analysis of interannual ET variability based on historical weather stations. Combining land surface process models assimilated with various observational data and other observations such as tree ring [22] records and runoff measurements could further enhance our understanding of the relationships between historical ET, hydrological climate changes, and human activities.

In conclusion, this study effectively reconstructs historical ET data from weather stations, revealing ET trends and climate change responses over the past 300 years. This helps to reassess the impact of historical climate change and human activities on the water cycle over several centuries and provides historical evidence for developing future climate adaptation strategies. Linking to the dynamics of other historical periods, such as socio-economic, will also help to deepen the understanding of how the 'human-water' system has changed over the centuries.

**Methods**

We trained a Random Forest model to estimate ET, using monthly latent heat (which can be linearly converted to ET) observations from FLUXNET2015 [8] flux stations as the target variable. The predictor variables included precipitation, temperature, aridity index, and rooting depth. In the training set, precipitation and temperature data came from FLUXNET2015 observations. The aridity index [23] was used to measure the long-term aridity variations of the stations across the globe. Rooting depth data, derived from previous global estimates [24], was used to represent the influence of vegetation around the sites. Areas with deeper rooting depths can access deeper soil water during droughts, maintaining higher transpiration. Most sites have relatively shallow rooting depths. The model achieved high accuracy with 10-fold cross-validation, with an $R^2$ of 0.66 and a KGE of 0.76, which is notable given the difficulty of obtaining these predictors for historical periods.

The historical meteorological data came from a previous study [7] that integrated station observation data from different sources to create monthly station records including temperature, precipitation, pressure, and wet day counts. However, the number of stations with both precipitation and temperature data was limited. To apply the ET model to as many historical weather stations as possible, we used nearby weather stations to fill in missing temperature or precipitation values. If a station record contained only temperature (or precipitation), the missing precipitation (or temperature) value was filled using data from nearby stations (with latitude and longitude differences less than 0.5 degrees). If more than two nearby stations met this criterion, their average was used. Ultimately, we predicted monthly ET data for 5267 stations. In subsequent annual-scale analyses, if a station had more than two missing months of data in a year, that year's data was excluded (which accounted for a small portion of the dataset). Annual values for precipitation and temperature were generated similarly.

**Data availability**

The flux data are from FLUXNET2015 (https://fluxnet.org/data/fluxnet2015-dataset/), and historical weather station data is from HCLIM (https://www.nature.com/articles/s41597-022-01919-w).

**Supplementary Figures**

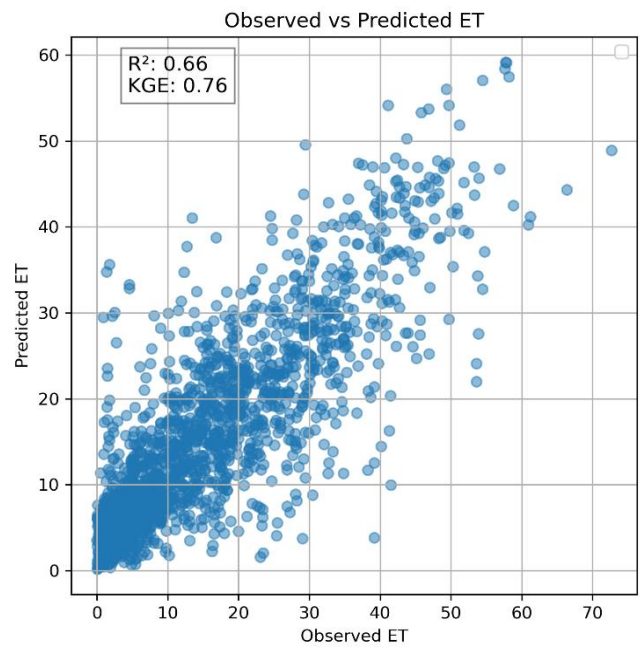

Supplementary Fig. 1 Accuracy of 10-fold cross-validation of the ET estimation model based on Random Forests.

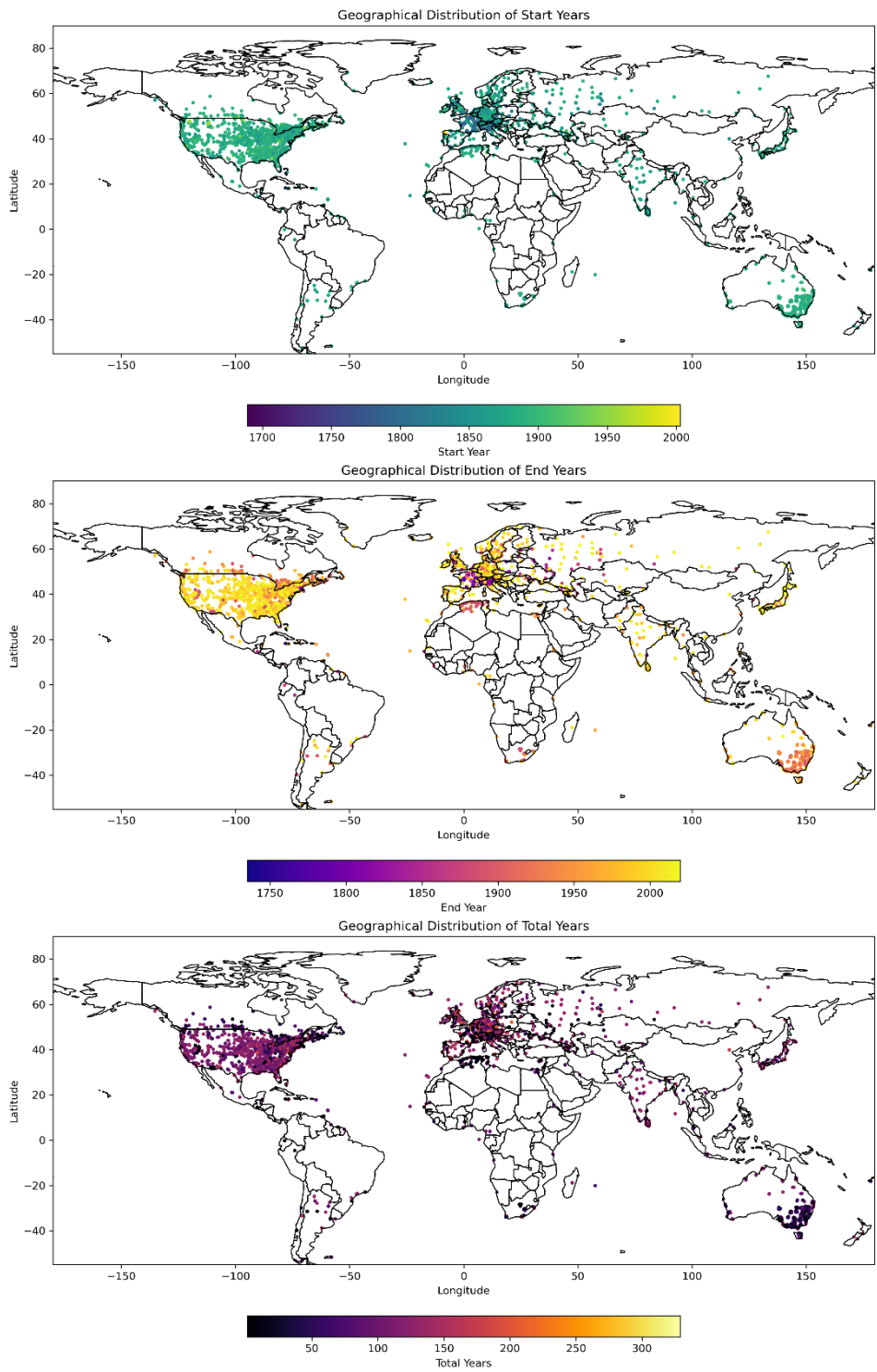

Supplementary Fig. 2 Geographic distribution of start year, end year and total number of years for the produced weather station ET data.

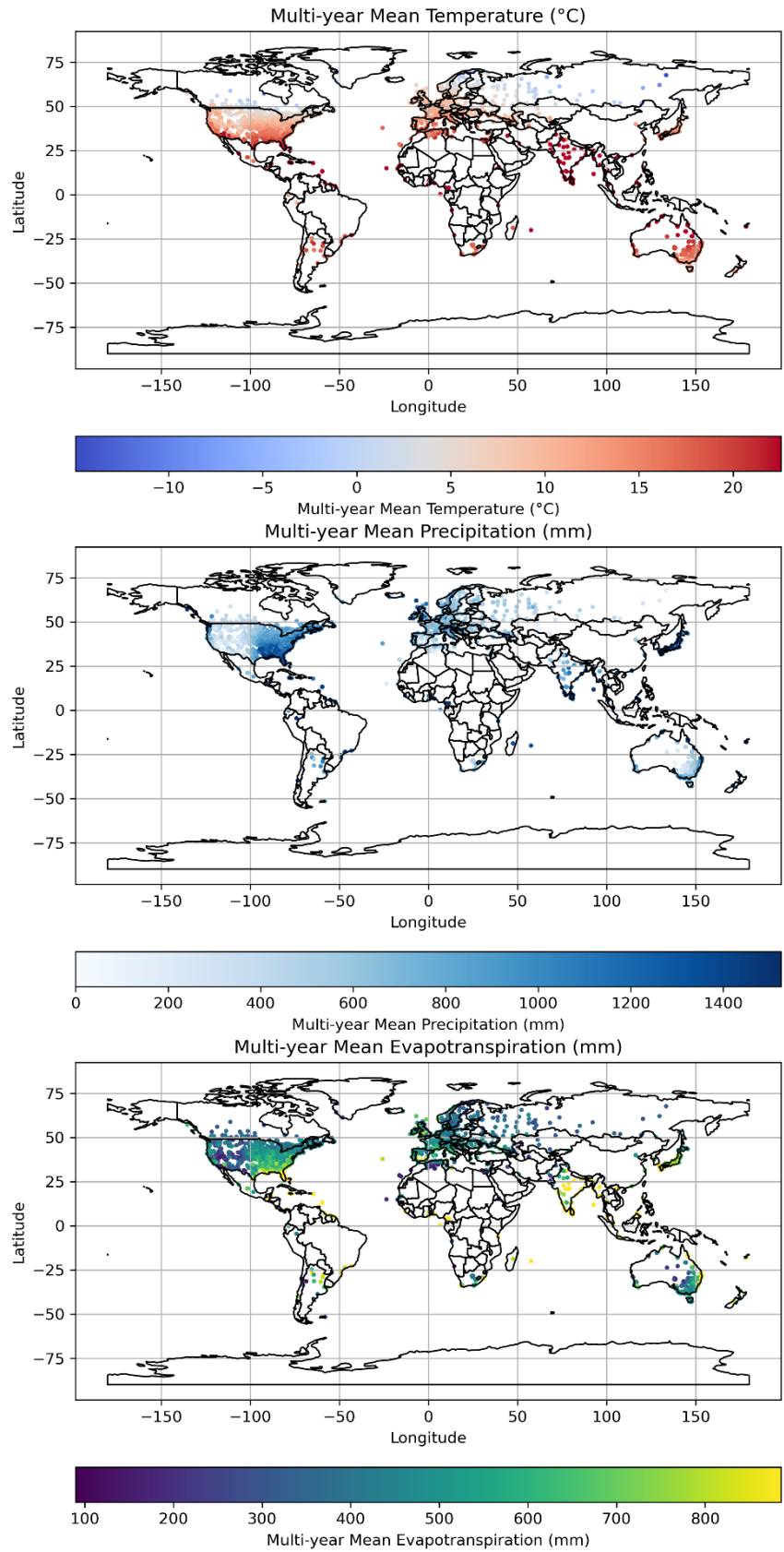

Supplementary Fig. 3 Multi-year averages of temperature, precipitation, and ET at stations with produced ET data.